\documentclass[a4paper]{jpconf}
\usepackage{graphicx}
\begin{document}
\title{Dynamics of the forward vortex cascade in two-dimensional quantum turbulence}

\author{Andrew Forrester and Gary A.\,Williams}
\address{Department of Physics and Astronomy, University of California, Los Angeles, CA 90095}

\ead{gaw@ucla.edu}

\begin{abstract}
The dynamics of the forward vortex cascade in 2D turbulence in a superfluid film is investigated using analytic techniques.  The cascade is formed by injecting  pairs with the same initial separation (the stirring scale) at a constant rate.  They move to smaller scales with constant current under the action of frictional forces, finally reaching the core size separation, where they annihilate and the energy is removed by a thermal bath.   On switching off the injection, the pair distribution first decays starting from the initial stirring scale, with the total vortex density decreasing linearly in time at a rate equal to the initial injection rate. As pairs at smaller scales decay, the vortex density then falls off as a power law, the same power law found in recent exact solutions of quenched 2D superfluids.\end{abstract}

Constant-current cascades have long played a role in understanding turbulence in fluids.  In classical 2D turbulence it is well known that there are two such cascades \cite{kraichnan80,boffeta12}, an inverse cascade of energy to large length scales with a $k^{-5/3}$ (Kolmogorov) energy spectrum, and a forward cascade of enstrophy (vorticity) to small length scales, with a $k^{-3}$ (or k$^{-3}$ln$(k)$) energy spectrum.  The dynamics of the free decay of such turbulence has been investigated with numerical simulations \cite{mcwilliams2000,chavanis2000,clercx2007,lacasce2008}, where the decay is dominated by the dynamics of the inverse cascade where 
like-sign regions of vorticity merge, heading toward the final state that is two macroscopic counter-rotating vortices.

The nature of the cascades in 2D quantum turbulence, with its discrete quantized vortices, is considerably less well understood.  A forward cascade of vortex pairs was found a number of years ago \cite{turb2001} for low vortex densities, comparable to or smaller than the density at the thermal Kosterlitz-Thouless transition.  More recently two numerical simulations of the Gross-Pitaevskii model at high vortex densities found instead a forward energy cascade with a $k^{-5/3}$ spectrum \cite{tsubota2010,chesler2013}.  Further complicating the picture is another set of simulations which found no direct cascade, but did find an inverse energy cascade with a roughly $k^{-5/3}$ spectrum \cite{anderson2013,anderson2014}. 

In this paper we show that analytic techniques can be used to study the dynamics of the growth and decay of the forward vortex cascade.  We find that the decay of the vortex density is initially linear in time, but then switches at longer times to a power-law decay with an exponent identical to that found in our recent exact solutions of quenched 2D superfluids \cite{forrester2013}.  We believe these are the first analytic solutions for the dynamics of any type of fluid turbulence.

We consider an incompressible superfluid film connected to a thermal bath at a temperature of 0.1 $T_{KT}$, where $T_{KT}$ is the critical Kosterlitz-Thouless temperature where thermally excited vortex pairs drive the superfluid density to zero.  The vortex dynamics are described by the same Fokker-Planck equation \cite{ahns} used in  describing the distribution of vortex pairs of separation $r$ \cite{forrester2013,chu2001}, with the addition of a 2D delta function to inject additional pairs of a fixed separation $R$ into the film at a rate $\alpha$,
\begin{equation}
\frac{{\partial \,\Gamma }}{{\partial \,t}} = \frac{1}{r}\,\:\frac{\partial }{{\partial r}}\left( {r\frac{{\partial \Gamma }}{{\,\partial r}} + 2\pi K\,\Gamma } \right) + \alpha \;{\delta ^2}(\vec r - \vec R)
\end{equation}
This equation is set in dimensionless form with lengths in units of the vortex core radius $a_0$, the vortex-pair distribution function $\Gamma$ in units $a_0^{-4}$, and the time in units of the vortex diffusion time, $\tau_0 = a_0^{2} / 2D$ with $D$ the vortex diffusion coefficient.  $K$ is the dimensionless superfluid density, $K = {{{\hbar ^2}{\sigma _s}}}/{{{m^2}{k_B}T}}$, and is determined from the Kosterlitz recursion relation \cite{kosterlitz}
\begin{equation}
\frac{{dK}}{{dr}} =  - 4{\pi ^3}{r^3}{K^2}\Gamma 
\end{equation}
with $\sigma_s$ the superfluid areal density and $m$ the $^4$He atomic mass.  The dimensionless injection rate $\alpha$ is given by 
$\alpha  = a_0^2\dot Q{\tau _0}$ where $\dot Q$ is the number of vortex pairs injected per unit area per time at random positions and orientations across the plane.

The steady-state solutions of Eq.\,1 where $\partial \Gamma /\partial t = 0$ were previously found in \cite{turb2001}, but only for $r < R$, while for dynamics the vortices at $r > R$ also need to be included.  In the limit of low injection rates the vortex densities are well below the densities at $T_{KT}$, and the superfluid fraction is unaffected by the vortices.  The solution for $r < R$ is then a constant value
\begin{equation}
\Gamma = {\Gamma _0} = \alpha /2{\pi}K_0 
\end{equation}
where $K_0$ is the value at $r =1$ where $\sigma_s$ equals the "bare" superfluid density $\sigma_s^0$.  
For $r > R$ the solution is a quasi-thermal distribution extending from $R$, $\Gamma =  \Gamma _0 \,(r -R)^{-2 \pi K_0}$, and which arises from injected pairs initially at separation $R$ getting a thermal kick to higher separation.  The steady-state flux of vortex pairs from Eq. 1 is 
$J =  - (r{{\partial \Gamma } \mathord{\left/
 {\vphantom {{\partial \Gamma } {\partial r + 2\pi K \Gamma)}}} \right.
 \kern-\nulldelimiterspace} {\partial r + 2\pi K \Gamma))}}$, and for this solution is a constant, $J = -\alpha$ for $r < R$, and zero for $r > R$.  

The two curves in Fig.\,1 for the lowest values of $\alpha$ show the steady-state distribution of pairs for the low-injection regime calculated directly from Eqs.\,1 and 2 using fixed-step Runge-Kutta iteration with the injection scale $R$ = 400, in agreement with the above form of Eq.\,3.  The delta function is approximated with a strongly peaked Gaussian form with spatial width of 2.  The superfluid fraction remains unity for $r < R$, though at $\alpha = 1\times10^{-12}$ the vortex density is getting high enough that there is a very slight reduction for $r > R$.
\begin{figure}[b]
\begin{minipage}{18pc}
\includegraphics[width=18pc]{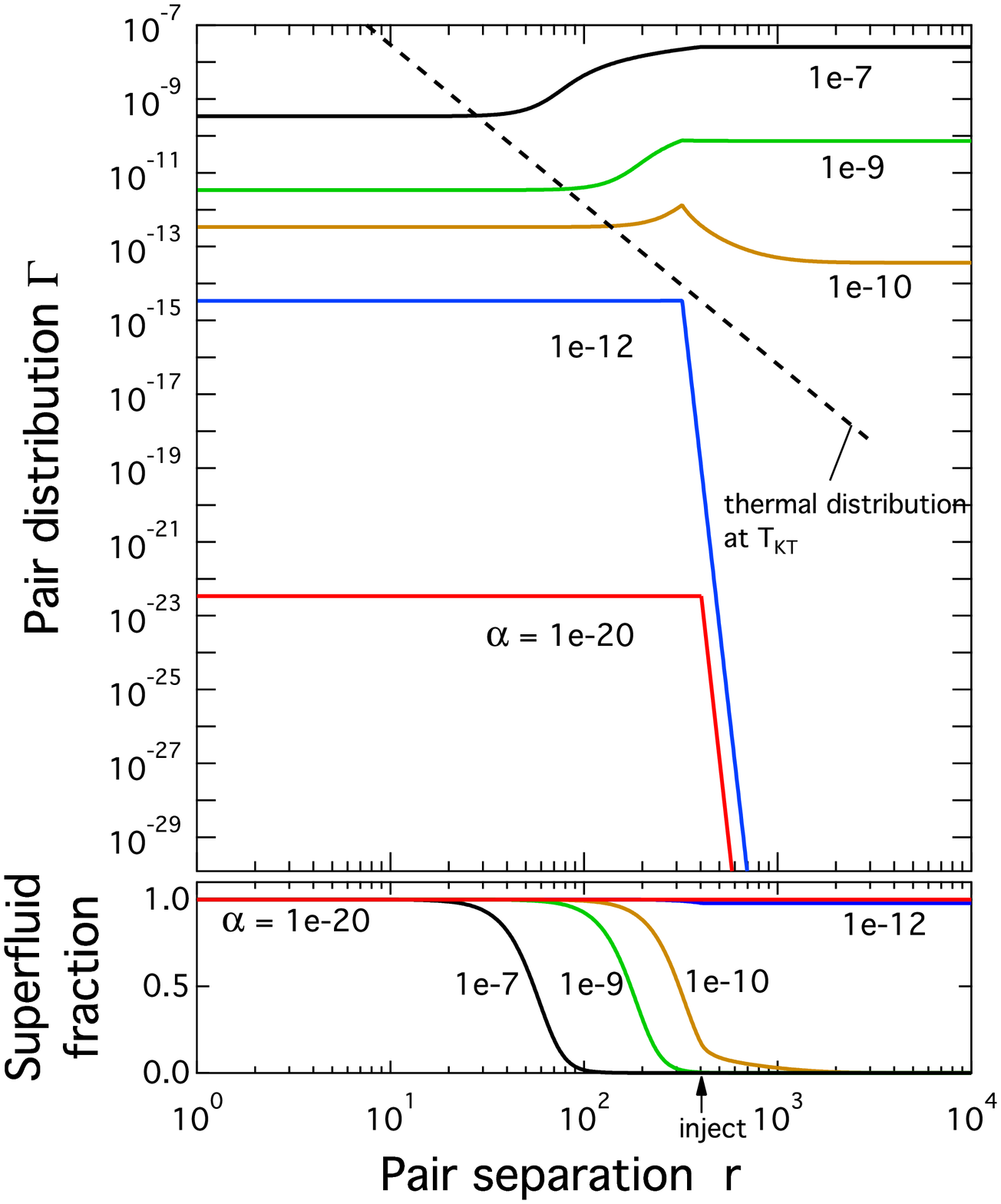}
\caption{\label{label}Steady-state solutions of Eqs.\,1 and 2 for the vortex-pair distribution function at different  injection rates $\alpha$. }
\end{minipage}\hspace{2pc}%
\begin{minipage}{18pc}
\includegraphics[width=18pc]{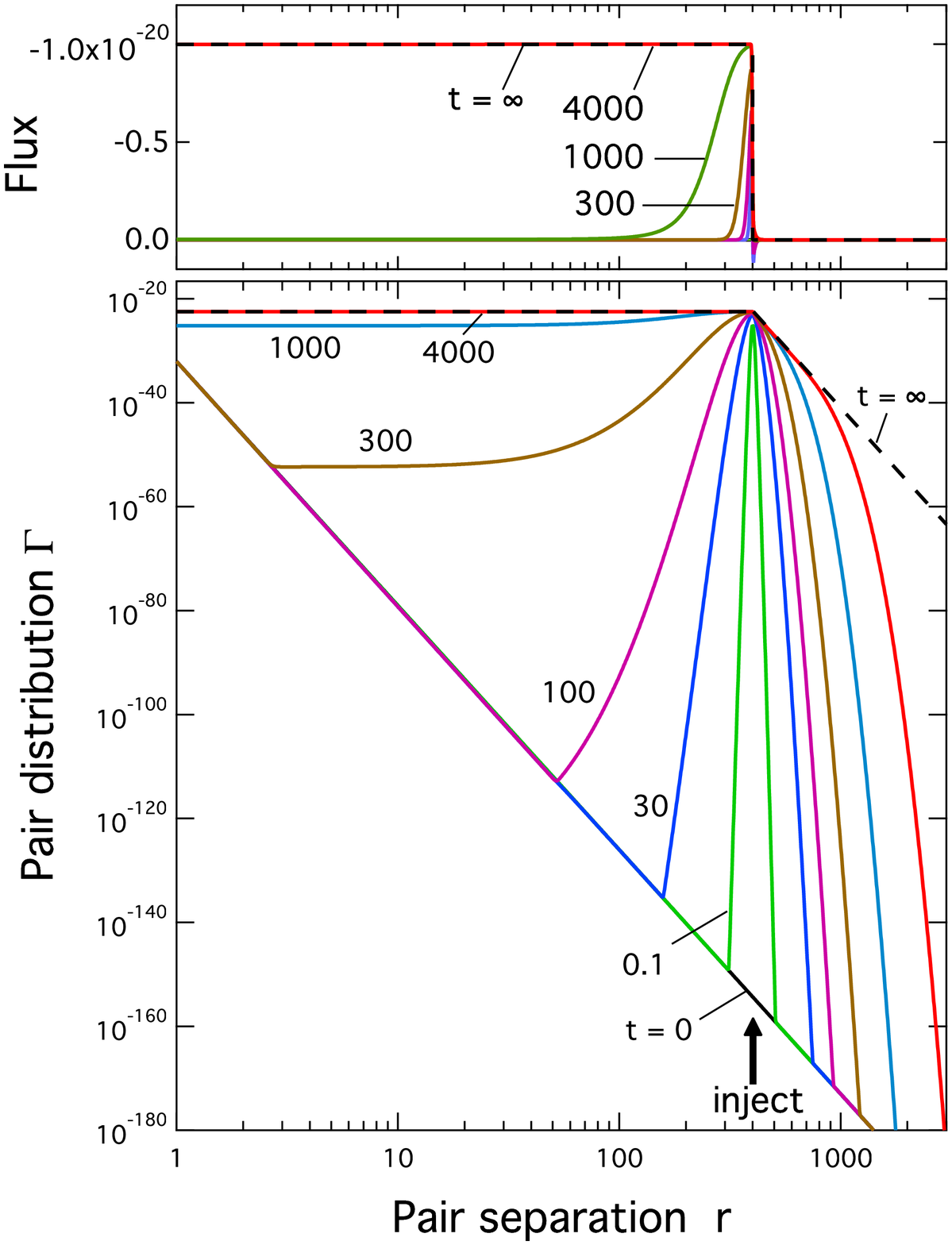}
\caption{\label{label}Growth of the pair distribution and current in time for $\alpha = 1\times 10^{-20}$.}
\end{minipage} 
\end{figure}

The upper two curves in Fig.\,1 show the distribution in the limit where the vortex density becomes comparable to that found at $T_{KT}$, indicated by the dashed curve in Fig.\,1.  The superfluid fraction is rapidly driven to zero at a finite length scale $r_0$ that depends on the injection rate.  This effect on the superfluid density shows that this turbulent state is not just characterized by isolated dipole pairs, but in fact is a complex many-body state of smaller pairs screening the interaction of larger pairs to drive down the superfluid density.  To a very good approximation the solutions in this regime can be represented as
\begin{eqnarray}
\begin{array}{l}
\Gamma  = {\Gamma _0}\quad \quad\quad\quad\quad\quad\quad(r < R)\\
\quad  = {\Gamma _0}(1 + \ln (r/{r_0}))\quad ({r_0} < r < R)\\
\quad  = {\Gamma _0}(1 + \ln (R/{r_0}))\quad (r > R).
\end{array}
\end{eqnarray}
The steady-state flux for this solution is also $J = -\alpha$ for $r < R$, and zero for $r > R$.

The time dependence of the cascade growth can be studied by solving Eqs.\,1 and 2 as a function of both time and distance using standard numerical techniques, starting from thermal equilibrium at 0.1 $T_{KT}$ and switching on the injection at $t = 0$.  Figure 2 shows the growth of the distribution and the pair current for the very low injection rate $\alpha = 1\times10^{-20}$.  Initially the distribution just broadens as the pairs get thermal kicks to larger and smaller separations, but the frictional forces on the vortex cores also give rise to a net current of pairs to smaller separations.  After a few hundred diffusion times the decaying pairs reach the core size separation where they annihilate, and the energy is pulled out by the thermal bath.  At longer times the distribution becomes constant at $r < R$ and increases uniformly toward the equilibrium value.  The current is initially only appreciable near the injection scale, and then finally at very long times reaches the value of $-\alpha$, the cascade state where the rate of pairs being injected at $R$ is equal to the rate of pairs being pulled out by the thermal bath at the scale $a_0$.  The total vortex density is shown in Fig.\,3, with initially a linear increase with slope $\alpha$ before any pairs are pulled out at $a_0$, and then a leveling off towards equilibrium once pairs begin annihilating.  
 \begin{figure}[t]
\begin{center}
\includegraphics[width=0.75\textwidth]{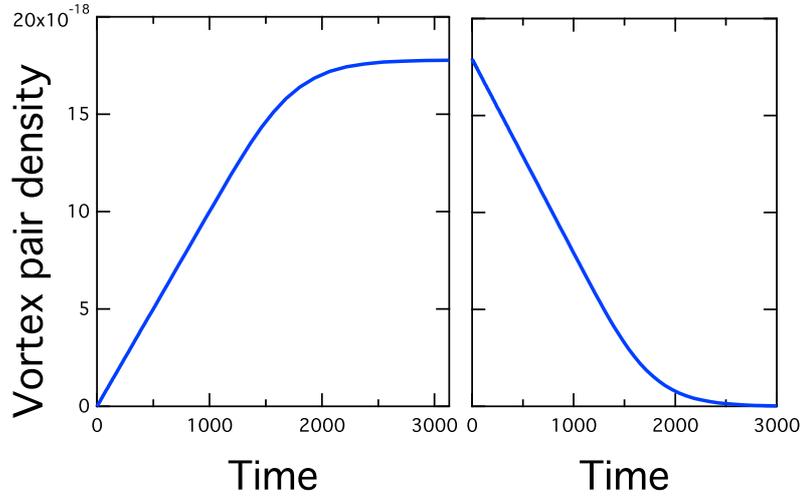} 
\end{center}
\caption{Growth (left) and decay (right) of the vortex-pair density, for injection rate $\alpha = 1\times10^{-20}$.}
\label{fig1}
\end{figure}
\begin{figure}[b]
\begin{minipage}{18pc}
\includegraphics[width=18pc]{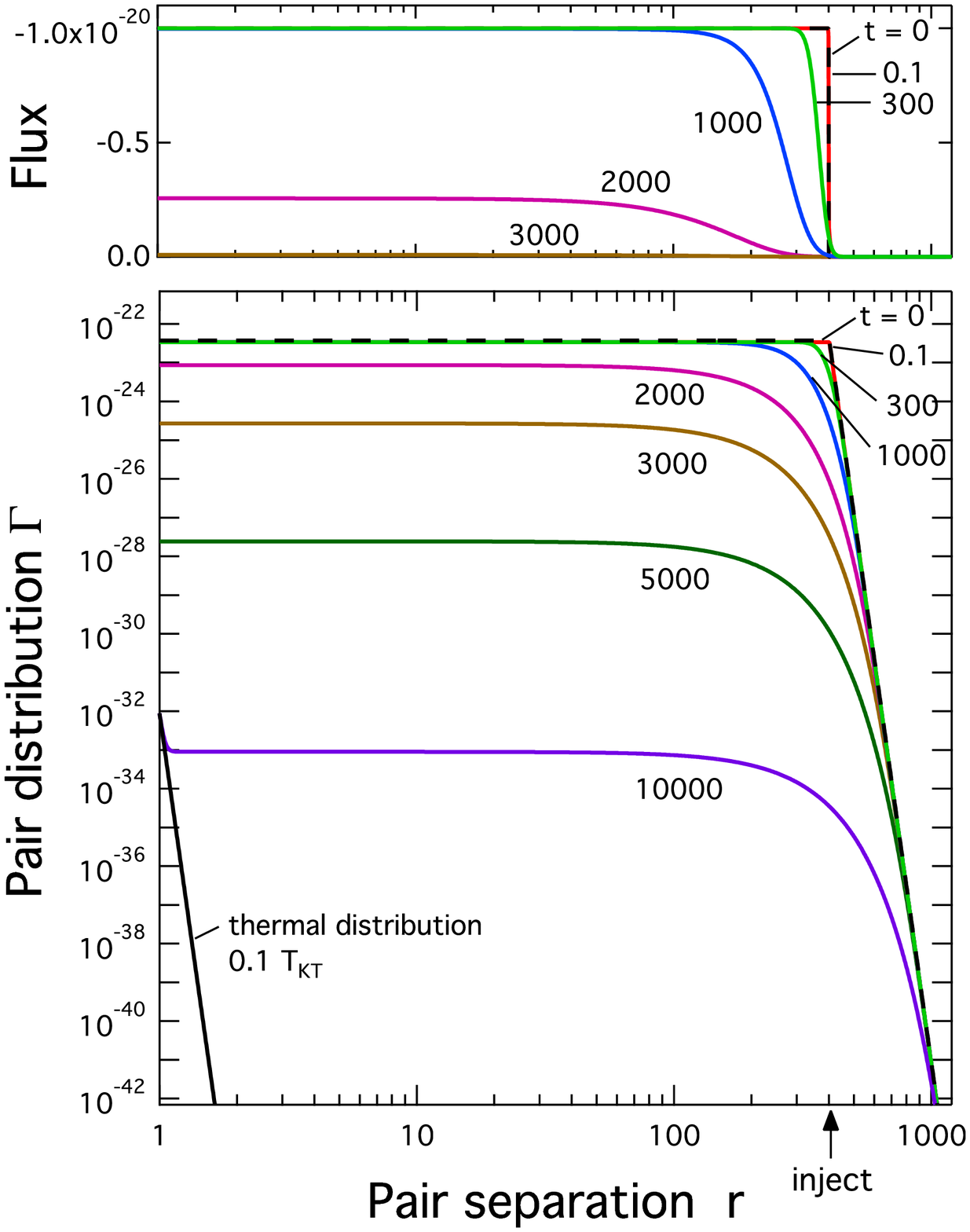}
\caption{\label{6}Decay of the pair distribution function and current after switching off the injection $\alpha = 1\times10^{-20}$. }
\end{minipage}\hspace{2pc}%
\begin{minipage}{20pc}
\includegraphics[width=20pc]{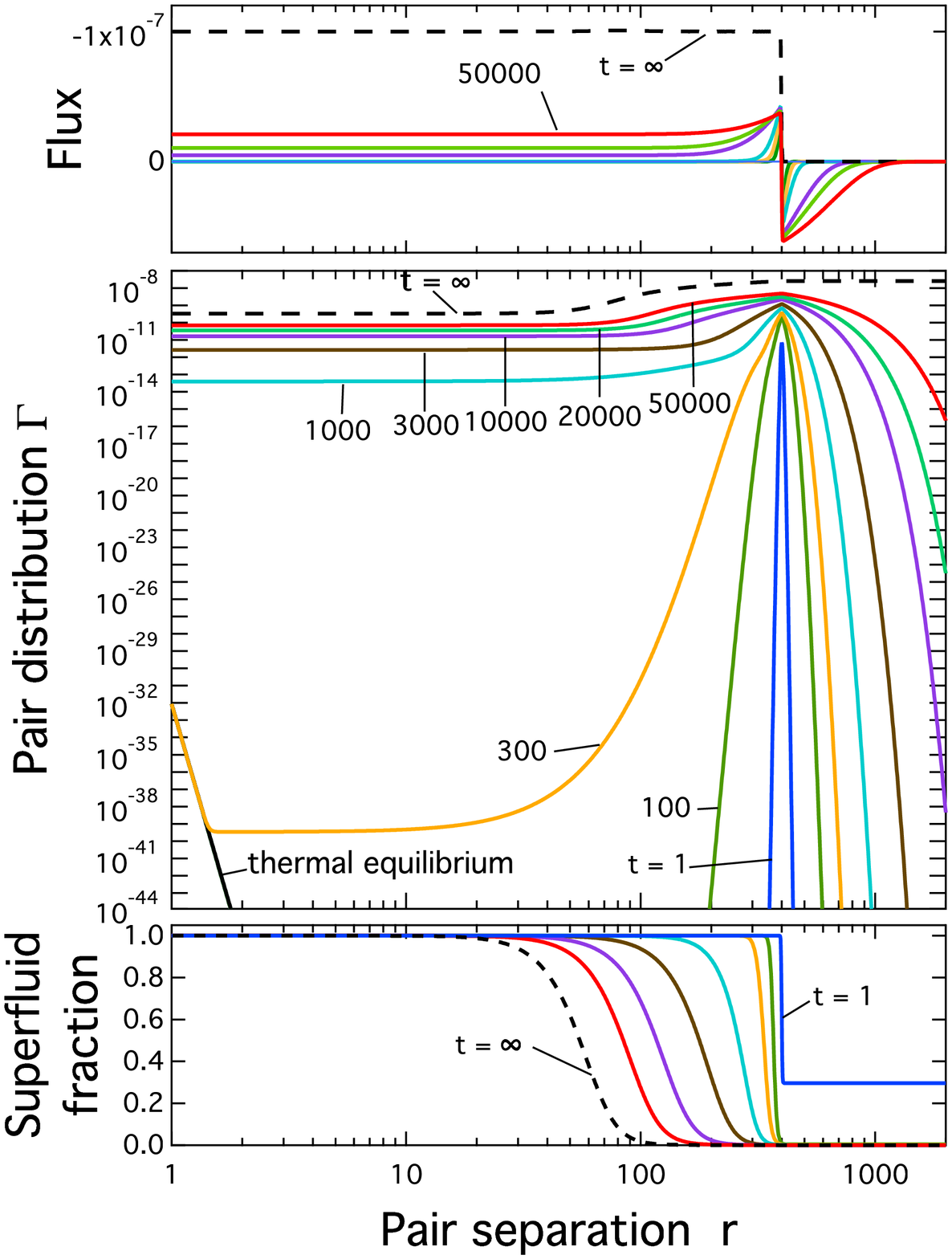}
\caption{\label{7}Growth of the pair distribution, current, and superfluid fraction for injection $\alpha = 1\times10^{-7}$.}
\end{minipage} 
\end{figure}

On switching off the injection, the decay of the cascade starts at the injection scale, as shown in Fig.\,4, since the current of pairs away from $R$ is no longer being replenished by the injected pairs.    Once the region of diminished pairs begins to reach the scale $a_0$ the distribution starts to uniformly decrease, and in fact in this regime the solution of Eqs.\,3 and 4 is the exact solution found in \cite{forrester2013} for a quenched 2D superfluid with an initial temperature equal to 0.1 $T_{KT}$.  At long times, however, it comes to equilibrium at the bath temperature.  The decaying vortex density is shown in Fig.\,3, initially a linear decay with slope $-\alpha$, since this is still the rate at which pairs annihilate, but then at long times falling off as $t^{-(\pi K_0 - 1)}$, the value predicted in \cite{forrester2013}.

In the limit of higher injection rates where the superfluid density is driven to zero at finite scales ($\alpha > 1\times10^{-11}$) the dynamics of the cascade becomes more complicated.  The gradient in the superfluid density $K$ in Eq.\,1 leads to an additional  current toward large scales that subtracts from the forward current and tends to slow down both the growth and decay of the cascade.  This is shown in Fig.\,5 for the growth of the cascade at $\alpha = 1\times10^{-7}$.   As the pair distribution grows the superfluid fraction is depressed to smaller and smaller length scales.  The current in the cascade region $r < R$ only grows slowly since the pairs being injected are divided, with part going in the forward direction and others increasing their separation in the inverse direction, driven by the gradient in $K$.  The vortex density in the cascade region grows more slowly than linear in time due to this division, with fits giving a variation as approximately $t^{2/3}$.

In this analysis we find no evidence of a forward energy cascade with spectrum $k^{-5/3}$ as found in the earlier simulations \cite{tsubota2010,chesler2013}, and this may be due to the relatively dilute vortex densities considered here, in which there will only be annihilation of same-pair vortices when their separation reaches $a_0$.  The vortex densities in Refs.\,\cite{tsubota2010} and \cite{chesler2013} were quite high, with vortices on a large fraction of the lattice sites, and in this situation there will be a much higher probability of annihilation of opposite-circulation vortices on different pairs, which might give rise to the observed forward energy cascade.  The densities are so high that these systems are certainly not superfluid at length scales much larger than the vortex core size.   A further difference from our system is that the Gross-Pitaevskii model that is simulated is a compressible superfluid, and the sound waves excited in the turbulence may also contribute to energy flows.  And finally, these  simulations had zero dissipation, whereas our forward cascade requires an arbitrarily small but finite dissipation to develop.  The dissipation only enters into the dynamics through the vortex diffusion constant that sets the time scale;  it does not affect the steady-state solutions of Eqs.\,3 and 4.  Weak dissipation still allows the possibility of "inertial" cascades, as shown by the simulations of the classical 2D Navier-Stokes equation with finite viscosity \cite{boffetta}.  

We also do not observe an inverse cascade of energy to large scales.  There is an initial outward current current as pairs build up in the region $r > R$, but this falls to zero in the steady state for both the low and high injection rates.  An inverse cascade requires a clustering of like-sign vortices in a negative-temperature state \cite{anderson2013,anderson2014}, but our Kosterlitz-Thouless model assumes from the start a uniform distribution of vortices with positive and negative circulation.

\section*{Acknowledgements}

This work was supported in part by the U.\,S.\, National Science Foundation, Grant No. DMR 0906467.

\section*{References}
\bibliography{Turbulence}
\end{document}